\documentclass[prl,preprint,amsmath,showpacs,groupedaddress,superscriptaddress]{revtex4}
\usepackage[english]{babel}
\usepackage[applemac]{inputenc}
\usepackage{latexsym}
\usepackage{hyperref}
\usepackage{graphicx}
\usepackage{epstopdf}
\usepackage{times} 

\begin{document}




\title {Magneto-optical behaviour of EuIn$_2$P$_2$}

\author{F. Pfuner}
\affiliation{Laboratorium f\"ur Festk\"orperphysik, ETH Z\"urich,
CH-8093 Z\"urich, Switzerland}

\author{L. Degiorgi}
\affiliation{Laboratorium f\"ur Festk\"orperphysik, ETH Z\"urich,
CH-8093 Z\"urich, Switzerland}

\author{H.~R. Ott}
\affiliation{Laboratorium f\"ur Festk\"orperphysik, ETH Z\"urich,
CH-8093 Z\"urich, Switzerland}

\author{A.~D. Bianchi}
\affiliation{Department of Physics, University of Irvine, Irvine CA 92697, U.S.A.}

\author{Z. Fisk}
\affiliation{Department of Physics, University of Irvine, Irvine CA 92697, U.S.A.}\

\date{\today}

\begin{abstract}
We report results of a magneto-optical investigation of the Zintl-phase compound EuIn$_2$P$_2$. The compound orders magnetically at $T_C$=24 K and exhibits concomitant large magnetoresistance effects. For $T\le$50 K and increasing magnetic fields we observe a transfer of spectral weight in $\sigma_1(\omega)$ from energies above 1 eV into the low-energy metallic component as well as into a mid-infrared signal centered at about 600 cm$^{-1}$. This latter absorption is reminiscent to what has been seen in a large variety of so-called Kondo materials and ascribed to excitations across the hybridization gap. The observed gain of Drude weight upon increasing magnetic field suggests an enhancement of the itinerant charge-carrier concentration due to the increasing magnetization, a phenomenon that was previously observed in other compounds which exhibit colossal magnetoresistive effects.
\end{abstract}
\pacs{75.47. Gk, 78.20.-e}
\maketitle

EuIn$_2$P$_2$ belongs to the broad class of materials known as Zintl compounds. The Zintl concept for the formation of compensated-valence intermetallic compounds is based on a complete charge transfer from an alkali or alkaline-earth element to a post-transition element of the groups 13 (IIIA) or 15 (VA) in the periodic table \cite{kauzlarich}. The concept was later extended to synthesize more complicated intermetallic compounds with new types of structures and a variety of physical properties, such as magnetic order and superconductivity \cite{holm,deakin,jiand,fisher}. Concerning magnetism, Zintl-phase compounds with rare-earth elements as one of the regular constituents were investigated. In a previous study, the onset of magnetic order in Yb$_{14}$MnSb$_{11}$ was observed \cite{fisher}. In this case, the rare-earth element Yb adopts a divalent and hence nonmagnetic 4f-electron configuration; the order is among the moments residing on the Mn ions. For EuIn$_2$P$_2$, the magnetic order involves the local moments of the partially occupied Eu 4f-electron orbitals and is thought to be induced by common exchange interactions. In the latter case, the onset of magnetic order is accompanied by a considerable reduction of the electrical resistance in the ordered state and related magnetoresistance effects upon the application of external magnetic fields \cite{jiand}. In view of similar observations in previous studies of materials, such as the perovskites $L$VO$_3$ ($L$=La, Lu and Y) \cite{nguyen}, EuO \cite{shapira} and EuB$_6$ \cite{guy}, and the discovery of giant magneto-optical effects, at least in EuB$_6$ \cite{broderick1,broderick2,caimi}, it seemed of interest to probe the optical properties of EuIn$_2$P$_2$  in a wide spectral range and, in addition, their variation with temperature and external magnetic field.

The magnetic susceptibility $\chi(T)$ of EuIn$_2$P$_2$ clearly reflects a magnetic transition at $T_C\sim$24 K \cite{jiand}. Above $T_C$, $\chi(T)$ is isotropic and exhibits a Curie-Weiss type behaviour with a paramagnetic Curie temperature $\Theta=27$ K. Below $T_C$, $\chi(T)$ depends on the orientation of the external magnetic field with respect to the crystal axes. The saturation magnetization is 7.04 $\mu_B$ per formula unit, very close to what is expected for a divalent configuration of the Eu ions. The results of the $\chi(T)$ and additional $M(H)$ data measured along different crystalline axes were interpreted as mirroring a canted magnetic order among ferromagnetically aligned Eu$^{2+}$ moments with an alternating tilting out of the basal plane of the hexagonal crystal structure \cite{jiand}.

As far as the dc transport is concerned, at $T\ge$100 K the magnitude of the electrical resistivity $\rho(T)$ of EuIn$_2$P$_2$ is of the order of m$\Omega$cm but the slope of its temperature dependence is positive. Below 100 K, $\rho$ increases with decreasing temperature and reaches a sharp maximum at $T_C$. The slope $\delta\rho/\delta T$ between 30 and 60 K was fitted with an exponential temperature dependence and claimed to be an indication of a gap of the order of 3 meV \cite{jiand}. The substantial drop of $\rho(T)$ below $T_C$, however, indicates the onset of metallic conductivity.

An intimate relation between the magnetization and the electronic conductivity, leading to large or colossal magnetoresistive effects (CMR), has been identified in a variety of compounds (e.g., EuB$_6$ \cite{guy,wigger} and the well-known manganites \cite{jin}), which were intensively studied both experimentally and theoretically. Most interpretations rest on magnetically induced shifts of band edges inducing substantial changes in the concentration of itinerant charge carriers \cite{caimi,wigger}. An obvious consequence of this scenario is the expectation of a redistribution of spectral weight between excitations at higher energies in the optical absorption spectrum and the Drude component \cite{pereira}. With a series of magneto-optical studies of Eu$_{1-x}$Ca$_x$B$_6$ it has been 
demonstrated that the enhancement of the Drude component is induced upon increasing the magnetization, either by varying the temperature and actuating a spontaneous magnetization in the ordered state or by increasing an externally applied magnetic field \cite{caimi}. Our present work reveals an analogous shift of spectral weight from high frequencies into the Drude-type absorption of the excitation spectrum upon the onset of magnetic order in EuIn$_2$P$_2$, thus confirming the general trend in the behaviour of this sort of materials exhibiting CMR effects.

Large single crystals of EuIn$_2$P$_2$ were grown in In flux. EuIn$_2$P$_2$ crystallizes with a hexagonal structure, the space group is P6(3)/mmc. The lattice consists of Eu layers in the $ab$ plane, separated by In$_2$P$_2$ layers alternating along the $c$-axis \cite{jiand}. Within the Zintl concept each Eu atom donates electrons to the In$_2$P$_2$ layers. It was shown \cite{jiand} that this particular structure of EuIn$_2$P$_2$ supports a total of 18 electrons for each (In$_2$P$_2$)$^{2-}$ slab. 

The optical reflectivity $R(\omega)$ of the title compound was measured in a broad spectral range from the far infrared (FIR) up to the ultraviolet, and as function of both temperature (1.6-300 K) and magnetic field (0-7 T) \cite{caimi}. The optical conductivity was extracted by a Kramers-Kronig transformation of $R(\omega)$. For this purpose the $R(\omega)$ spectra were extrapolated towards high frequencies (i.e., $\omega\ge$70000 cm$^{-1}$) with $R(\omega)\sim\omega^{-s}$, 2$\le s \le$4. The Hagen-Rubens (HR) ($R(\omega)=1-2\sqrt{\omega/\sigma_{dc}}$) extrapolation to $\omega\rightarrow0$ was employed at low frequencies (i.e., $\omega\le$ 50 cm$^{-1}$). The $\sigma_{dc}$ values in the HR extrapolation that are compatible with our $R(\omega)$ data were found to be in fair agreement with values from the dc transport experiments \cite{jiand}. This is shown in the inset of Fig. 1c, where we compare the magnetic field dependence of the $\sigma_{dc}$ values used in the HR extrapolation with those that were obtained from dc transport experiments, both at 30 K. Further details pertaining to the experimental technique and data analysis can be found elsewhere \cite{wooten,dressel}.

Figure 1 displays the measured $R(\omega)$ as a function of temperature (panel (a)), and as a function of magnetic field at 10 and 30 K (panels (b) and (c)), respectively. The main panels of Fig. 1 emphasize the far-infrared (FIR) range, where the temperature and magnetic field dependences are significant. Above 50 K the magnetic field dependence is negligible. The inset of Fig. 1a displays $R(\omega)$ across the entire measured spectral range at room temperature. It exhibits a metallic component with a plasma edge at very low frequencies ($\omega\leq$250 cm$^{-1}$), a strong infrared active phonon mode, peaking at 270 cm$^{-1}$ with a shoulder on its high frequency tail, a broad excitation centered at 600 cm$^{-1}$ and several absorptions above 1000 cm$^{-1}$. For each combination of temperature and magnetic field the high-frequency parts of all spectra merge.  Remarkable is the extremely low plasma edge which, as will be discussed in detail below, already suggests that EuIn$_2$P$_2$ is a system with a low itinerant charge-carrier density. The reflectivity is significantly influenced by magnetic field (Fig. 1b and 1c), much less so, however, by temperature (Fig. 1a).

The main panel of figure 2 presents the field dependence of $\sigma_1(\omega)$ at 30 K and at frequencies in the FIR range, while its inset emphasizes $\sigma_1(\omega)$ across the entire covered spectral range. Besides the sharp IR phonon mode and the effective metallic component, $\sigma_1(\omega)$ also reveals a strong absorption peaking at about 600 cm$^{-1}$. The first qualitatively important observation is that the metallic (Drude) component as well as the feature at 600 cm$^{-1}$ increase with magnetic field, i.e., they both gain spectral weight. The trend of the magnetic field dependence of $\sigma_1(\omega)$ shown for 30 K in Fig. 2 is representative for all $T\le$ 50 K. The enhancement of these respective mode strengths with magnetic field is only marginal in the $\sigma_1(\omega)$ spectra evaluated at temperatures above 50 K. 

The strong absorption centered at 600 cm$^{-1}$ bears a striking similarity with what has been often observed in heavy-electron or Kondo materials and was ascribed to excitations across the hybridization gap (hg) \cite{degiorgirmp,anders}. The features of the split Kondo resonance \cite{zawadowski} are indeed reflected in the shape of $\sigma_1(\omega)$. The hybridization between localized electron orbitals and the narrow valence (conduction) band in Kondo lattices leads to a splitting of that band. If the Fermi energy lies in the narrow band region, then heavy quasiparticles are formed. At low temperatures one expects, besides a coherent Drude peak in $\sigma_1(\omega)$ resulting from the renormalized heavy quasiparticles, an additional mid-infrared signal due to transitions across the so-called hybridization gap. The persistence of this signal up to room temperature suggests a rather high characteristic Kondo temperature of EuIn$_2$P$_2$. The already mentioned low density of itinerant charge carriers places the plasma edge of $R(\omega)$ at much lower frequencies than the hg excitation. Therefore, with increasing temperature the thermal broadening of the plasma edge is not sufficient to mask the hg signal, which is still visible at 300 K. At this point we note that a similar feature, also ascribed to excitations across a hybridization gap and clearly persisting up to room temperature, was recently observed around 200 cm$^{-1}$ in the Zintl compound, Yb$_{14}$MnSb$_{11}$ \cite{burch}. Finally the peaks in $\sigma_1(\omega)$ above 1000 cm$^{-1}$ (inset Fig. 2) are ascribed to electronic interband transitions, mostly involving the f and d orbitals of Eu, as well as the p and d orbitals of the post-transition element. A detailed assignment of these transitions will have to be based on results of a detailed calculation of the electronic band structure.

The Lorentz-Drude model, based on the classical dispersion theory for condensed matter, is the most common phenomenological approach in order to account for some of the features in the excitation spectrum \cite{wooten,dressel}. Besides the Drude term for the free charge-carriers' contribution, we consider several Lorentz harmonic oscillators (h.o.), across the covered spectral range. As an example, we present the components of the fit in zero magnetic field in Fig. 2. The effective metallic component cannot be represented by a single Drude term and a Lorentz h.o. at 150 cm$^{-1}$ must be added in order to account for part of its high frequency tail. Two h.o.'s are needed in order to reproduce the sharp phonon mode at 270 cm$^{-1}$ and its high-frequency shoulder at 310 cm$^{-1}$. Above 1 T the hybridization-gap feature is better described by two h.o.'s at 590 and 600 cm$^{-1}$, the first one being vanishingly small in zero field. It turns out that only the components peaking below 1000 cm$^{-1}$ (shown in the main panel of Fig. 2) vary significantly with temperature and magnetic field.

Although this phenomenological approach is rather simple, it allows for a quantitative evaluation of the magnetic field dependences of the Drude parameters $\omega_p$ and $\Gamma$, which determine the charge transport in EuIn$_2$P$_2$ and therefore also the CMR features. As pointed out above, the plasma frequency $\omega_p$ adopts extremely low values, even smaller than in Eu$_{1-x}$Ca$_x$B$_6$ \cite{caimi}, where $n_c$, the concentration of conduction electrons, is of the order of 10$^{-4}$ per unit cell \cite{wigger04}. Thus, EuIn$_2$P$_2$ must indeed be regarded as a system with a low concentration of itinerant charge carriers. The plasma frequency increases with increasing magnetic field at all temperatures below 50 K. The saturation to a constant value is reached at fields which increase with increasing temperature (Fig. 3a). This suggests a gain of the Drude spectral weight upon growing magnetization of the material. The Drude scattering rate does not vary with field above 30 K. It is considerably reduced above 1 T at 10 K (Fig. 3b), however, indicating a reduction of the scattering rate at low temperatures and high magnetic fields, i.e., in the saturated magnetically ordered state. The already mentioned additional h.o. at about 150 cm$^{-1}$, needed to account for the high-frequency tail of the effective metallic contribution, might be thought to be due to the band gap that is indicated by the dc transport data \cite{jiand}. However, its resonance energy is a factor of five larger than the previously claimed band gap, making this interpretation rather unlikely. More likely, this latter term in our fit procedure reflects excitations involving charges in quasi-localized states (e.g., with a large effective mass).  

The phenomenological decomposition of the optical conductivity into all its components also provides the basis for discussing the distribution of the spectral weight ($S$) in the absorption spectrum. Figure 4 shows the spectral weight distribution for some of these components as a function of magnetic field at selected temperatures. Included are the effective metallic component, i.e., the Drude term plus the h.o. at 150 cm$^{-1}$, the phonon signal at 270 cm$^{-1}$ together with its high frequency shoulder at 310 cm$^{-1}$ and the broad absorption peaking at 600 cm$^{-1}$. First, we note that the total mode strength of the phonon feature remains constant as a function of magnetic field at each temperature. The width (damping) of the lattice mode decreases at low temperatures. The resulting narrowing of the two h.o.'s for the phonon mode, combined with the gain of spectral weight of the Drude term in higher fields, leads to a suitable zero crossing of the dispersive part of the complex dielectric function with the frequency axis. Such a favorable interplay between the Drude and the h.o. components enhances the visibility of the high frequency shoulder of the phonon mode in $R(\omega)$ with increasing magnetic field at low temperatures. The appearance of the high frequency shoulder growing out of the main lattice mode possibly indicates the presence of a so-called localized phonon mode \cite{barker}. 

The most significant result of the spectral weight analysis presented in Fig. 4 is the remarkable gain of spectral weight of the effective metallic component (i.e., for frequencies $\omega\le$ 200 cm$^{-1}$) and the hybridization-gap feature centered at about 600 cm$^{-1}$. It is particularly pronounced at 30 K and even more so at 10 K (Fig. 4a and 4b). The renormalized density of states is significantly enhanced both at low temperatures and in high magnetic fields and is due to the increasing hybridization between the localized 4f states and the itinerant states of the d bands. This immediately leads to an enhancement of the transition probability, across the hybridization gap, i.e., a gain of the mode strength. The enhancements of the metallic component and hybridization-gap excitation tend to saturate in high magnetic fields at all temperatures. This saturation is actually expected because of the fully ordered magnetic state in high fields. The spectral weight for all components of $\sigma_1(\omega)$ remains almost constant at any fields above 50 K (Fig. 4c), confirming the direct experimental observation that at these temperatures, there is nearly no magnetic field dependence in the $R(\omega)$ spectra. In order to satisfy the optical sum rule, the gain of spectral weight of the low energy excitations of EuIn$_2$P$_2$ below 30 K is accounted for by an equivalent loss of signal at high energies, possibly above 1 eV. The transfer of weight from energies even higher than the experimentally covered spectral range seems to be a common feature of highly correlated systems \cite{degiorgirmp,kotliar}. It remains to be seen whether also in EuIn$_2$P$_2$, the spectral weight transfer into the low frequency range of $\sigma_1(\omega)$ is primarily driven by correlation effects.

The enhancement of the (Drude) metallic weight with magnetic field bears some similarities with our previous findings for the Eu$_{1-x}$Ca$_x$B$_6$ series \cite{caimi} and signals the release of localized charge carriers in the spin-polarized state of EuIn$_2$P$_2$. A variety of models was invoked to address the wealth of physical properties in CMR systems \cite{wigger,pereira,ghosh,lin,calderon}. Nonetheless, because of the similarity encountered in the magnetic field and temperature dependence of $\sigma_1(\omega)$ for both EuB$_6$ and EuIn$_2$P$_2$, it is worth recalling the scenario recently introduced by Pereira \textit{et al.}  \cite{pereira}. They proposed that the ferromagnetism of a low charge carriers system can be described within a double-exchange model. This theoretical approach can be regarded as an effective Kondo lattice problem in the limit of very small number of carriers. Pereira's model was successful in explaining several properties of ferromagnetic EuB$_6$ and particularly the evolution of the CMR effect and of the magneto-optical response in the Eu$_{1-x}$Ca$_x$B$_6$ series \cite{caimi}. The reduced itinerant carrier concentration places the Fermi level near a magnetization dependent mobility edge, which emerges in the spectral density because of the disordered spin background. Within this scenario it may then be argued that for both EuB$_6$ and EuIn$_2$P$_2$ the increasing magnetization due to increasing magnetic fields shifts the mobility edge away from the Fermi energy towards the lower band edge. This releases additional charge carriers and the Drude weight grows upon magnetizing the system.

In summary, we have presented magneto-optical data on the rare-earth based Zintl EuIn$_2$P$_2$ compound, which displays remarkable temperature and magnetic field dependence of the optical spectra at $T\le$ 50 K. We identified a non-negligible enhancement of spectral weight in $\sigma_1(\omega)$ into the low-energy metallic component with increasing magnetic field. This implies a release of additional charge carriers upon spin polarizing the material. Therefore, EuIn$_2$P$_2$ shares common features with other CMR compounds. In interpreting the data we favored the recently developed concepts based on the double-exchange scenario, which, although by no means unique, catches some of the experimental findings. Our results should motivate further theoretical and experimental work in order to strengthen our knowledge on the physical properties of EuIn$_2$P$_2$ and broaden our perspectives on CMR systems.

\begin{figure}[t]
  \begin{center}
		\includegraphics[width=0.75\columnwidth]{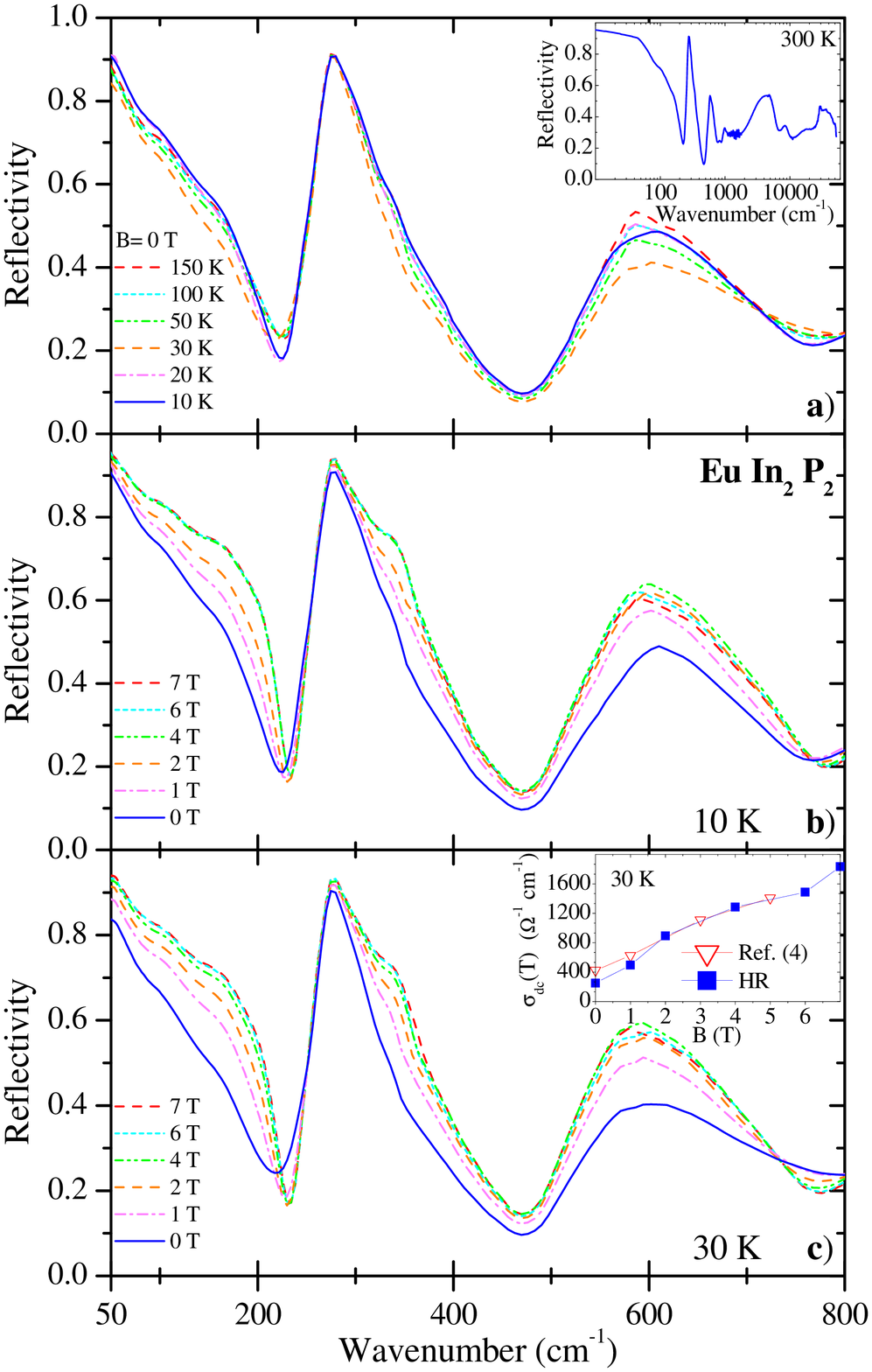}
		\caption{(color online) Reflectivity spectra $R(\omega)$ of EuIn$_2$P$_2$ in the infrared spectral range as a function of temperature at $B=0$ T (a). Panels (b) and (c) show $R(\omega)$ at 10 and 30 K, and in magnetic fields between 0~T and 7~T, respectively. The inset in panel (a) displays $R(\omega)$ at 300 K in the whole measured spectral range (logarithmic energy scale). The inset in panel (c) shows the comparison between the $\sigma_{dc}$ values inserted in the HR extrapolation and the dc data \cite{jiand} at 30 K and as a function of magnetic field.}
		\label{fig1}
	\end{center}
\end{figure}

\begin{figure}[t]
	\begin{center}
  	\includegraphics[width=.75\columnwidth]{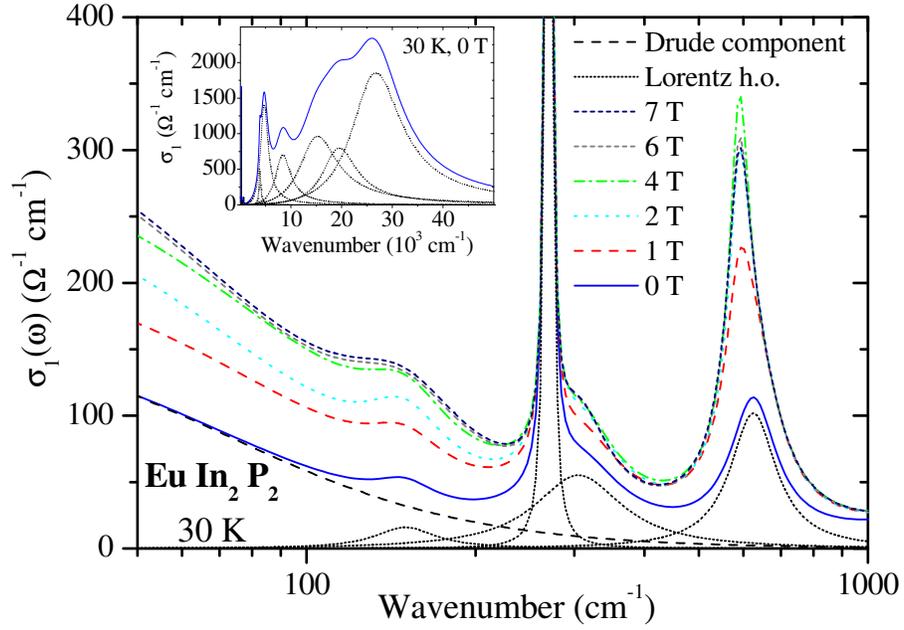}
		\caption{(color online) Real part $\sigma_1(\omega)$ of the low-frequency optical conductivity of EuIn$_2$P$_2$ at 30~K and as a function of magnetic field on a logarithmic energy scale. The Lorentz-Drude fit components (broken and dotted lines) are displayed for the 0 T spectra. The inset emphasizes $\sigma_1(\omega)$ and its Lorentz components in the high-frequency spectral range (linear energy scale).}
		\label{fig2}
	\end{center}
\end{figure}

\begin{figure}[t]
	\begin{center}
		\includegraphics[width=.75\columnwidth]{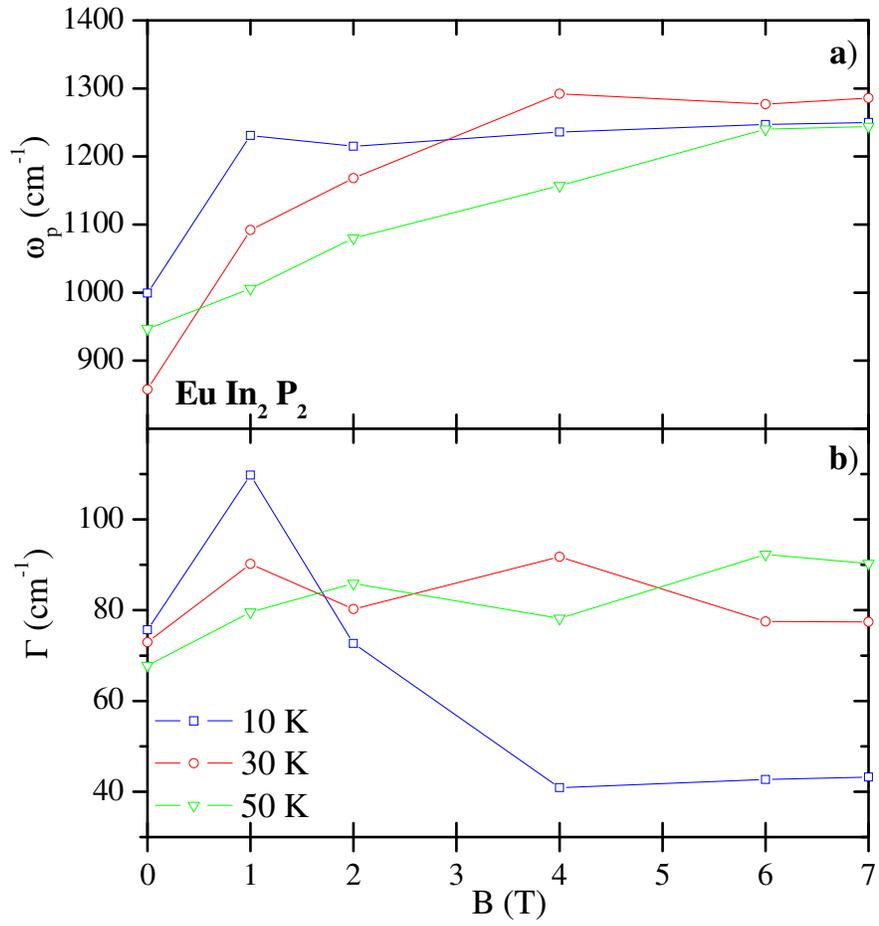}
		\caption{(color online) (a) Plasma frequency $\omega_{p}$ and (b) Drude scattering rate $\Gamma$, as a function of magnetic field at 10, 30 and 50 K.}
		\label{fig3}
	\end{center}
\end{figure}

\begin{figure}[t]
	\begin{center}
		\includegraphics[width=.75\columnwidth]{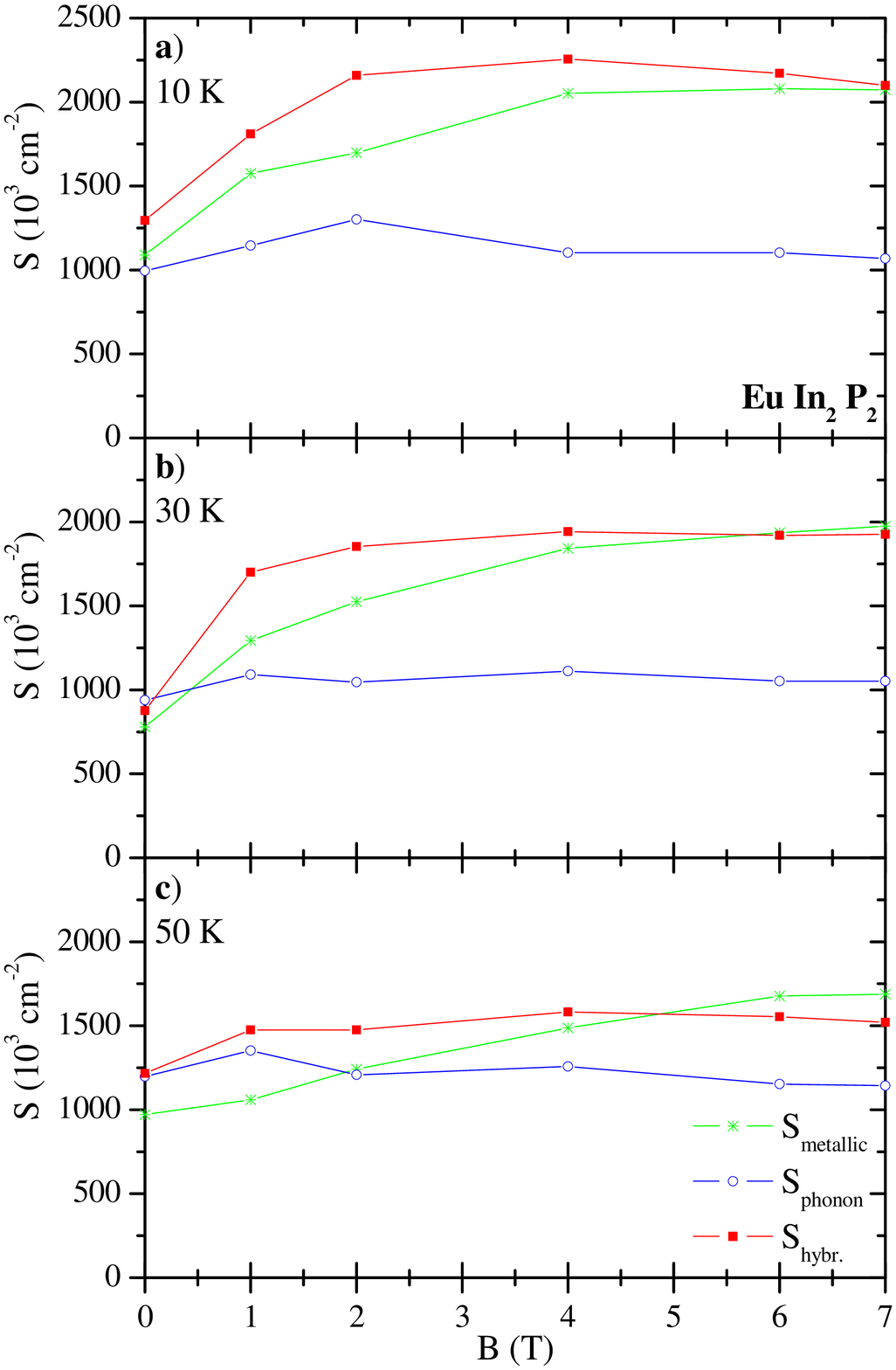}
		\caption{(color online) Magnetic field dependence of the spectral weight distribution at (a) 10, (b) 30 and (c) 50 K in the effective metallic component (S$_{metallic}$=$\omega_p^2$(Drude)+$\omega_p^2$(150 cm$^{-1}$)), the lattice phonon mode (S$_{phonon}$=$\omega_p^2$(270 cm$^{-1}$)+$\omega_p^2$(310 cm$^{-1}$)) and the hybridization gap feature (S$_{hybr.}$=$\omega_p^2$(600 cm$^{-1}$)+$\omega_p^2$(590 cm$^{-1}$)). $\omega_p^2$ is the squared plasma frequency and mode strength for the Drude term and h.o. at the resonance frequency given in the brackets, respectively.}
		\label{fig3}
	\end{center}
\end{figure}

\acknowledgments
The authors wish to thank J. M\"uller for technical help.
This work has been supported by the Swiss National Foundation for the Scientific Research, within the NCCR research pool MaNEP. Work at UC Irvine benefited from financial support of the US National Science Foundation under contract DMR-0433560.

\bibliographystyle{prsty}

\end{document}